\documentclass[11pt]{article}
\usepackage[utf8]{inputenc}
\usepackage{subcaption}

\usepackage{natbib}
\usepackage{graphicx}
\usepackage{amssymb}
\usepackage{amsfonts}
\usepackage{amsmath}

\newcommand{\LL}{\mathcal{L}}
\newcommand{\dif}{\mathrm{d}}

\DeclareMathOperator{\prominence}{\mathsf{prominence}}
\DeclareMathOperator{\maxep}{\mathsf{MXEP}}
\DeclareMathOperator{\stc}{\mathsf{STC}}
\DeclareMathOperator{\extrema}{\mathsf{extrema}}
\DeclareMathOperator*{\argmin}{arg\,min}

\newcommand{\prominenceof}[1]{\prominence\left(#1\right)}
\newcommand{\maxepof}[1]{\maxep\left(#1\right)}
\newcommand{\extremaof}[1]{\extrema\left(#1\right)}

\title{Estimating an Executive Summary of a Time Series: The Tendency}

\author{
Caio Alves\thanks{CSMD Division, Oak Ridge National Laboratory, Oak Ridge TN USA, 37831},
Juan M. Restrepo$\textsuperscript{*}$\thanks{Department of Mathematics University of Tennessee, Knoxville,  Knoxville TN USA, 37920},
Jorge M. Ramirez\textsuperscript{*}
\footnote{Notice: This manuscript has been authored by UT-Battelle, LLC, under contract DE-AC05-00OR22725 with the US Department of Energy (DOE). The US government retains and the publisher, by accepting the article for publication, acknowledges that the US government retains a nonexclusive, paid-up, irrevocable, worldwide license to publish or reproduce the published form of this manuscript, or allow others to do so, for US government purposes. DOE will provide public access to these results of federally sponsored research in accordance with the DOE Public Access Plan (https://www.energy.gov/doe-public-access-plan). 
}}

\begin{document}

\maketitle

\begin{abstract}
In this paper we revisit the problem of decomposing a signal into a tendency and a residual. The tendency describes an executive summary of a signal that encapsulates its notable characteristics while disregarding seemingly random, less interesting aspects. Building upon the Intrinsic Time Decomposition (ITD) and information-theoretical analysis, we introduce two alternative procedures for selecting the tendency from the ITD baselines. The first is based on the maximum extrema prominence, namely the maximum difference between extrema within each baseline. Specifically this method selects the tendency as the baseline from which an ITD step would produce the largest decline of the maximum prominence. The second method uses the rotations from the ITD and selects the tendency as the last baseline for which the associated rotation is statistically stationary.  We delve into a comparative analysis of the information content and interpretability of the tendencies obtained by our proposed methods and those obtained through conventional low-pass filtering schemes, particularly the Hodrik-Prescott (HP) filter. Our findings underscore a fundamental distinction in the nature and interpretability of these tendencies, highlighting their context-dependent utility with emphasis in multi-scale signals. Through a series of real-world applications, we demonstrate the computational robustness and practical utility of our proposed tendencies, emphasizing their adaptability and relevance in diverse time series contexts. 

	% \noindent
	% \emph{Keywords and phrases.}
	%    time series analysis, trends, diffusion map filter, surrogate models.

	% \noindent
	%   MSC 2020: 62M10; 60G35.
\end{abstract}

% \begin{keywords}
% 	   Time series analysis, trends, diffusion map filter, surrogate models.
% \end{keywords}

\section{Introduction}

A common task in time series analysis is to find structural, correlational or causative, relations between the time series structure and the mechanisms that generated the time series in the first place. \cite{itd} proposes a time series denoted as {\it the tendency} of a signal, which can best be described as an executive summary of the original signal. 

What we are seeking, qualitatively,  from this executive summary 
is that it captures the notable characteristics  of the original signal, leaving out aspects of the signal that are of little interest because of their apparently random behavior.
Further, the complexity of the tendency should be lower than that of the signal itself. What, precisely, we mean by notable and ignorable is the key challenge in formulating a tendency. A first attempt at defining a tendency appears in \cite{itd}
and exploits a specific fully adaptive time series decomposition of the time series, and then uses a criteria to assign to one of its components the role of the tendency. The decomposition we used is the Intrinsic Time Decomposition (ITD hereon) proposed in \cite{frei-osorio06} and summarized here in Section \ref{sec:background}. The ITD decomposition was defined originally in purely algorithmic terms, and in \cite{itd} it was found to be the result of applying iteratively a diffusion map to the original signal. The criterion used to assign one of the components of the ITD as the  tendency resulted from applying an information-theoretical analysis. However, the criterion produced tentative results unless  the time series  in question had a  significant length to populate stably an empirical histogram, limiting its applicability. In this study we revisit the definition of the tendency, whilst retaining the diffusion filter decomposition. In Section \ref{sec:tendency} we propose a couple of alternative tendencies, equipped with a significantly more robust criteria for their selection. Further, we provide a clearer 
definition of what we want to keep in the executive summary. 

The {\it tendency} $\{T(i) \}_{i=1}^N$ is a generalization of a trend, for an $N$-point  time series $\{Y(i)\}_{i=1}^N$, where $i$ is the sequential index. In this study we propose the tendency in two ways. Because we are using two different criteria, these tendencies will have different interpretations. They will be introduced in Section \ref{sec:tendency}. To endow the tendency with interpretability we examine the residual, defined as $r(i) = Y(i)-T(i)$, for $i=1,2,..,N$. In Section \ref{sec:interpretation} we contrast the information content of the residual, to that obtained by a popular low-pass filtering scheme, which is in use in finance and economics for the purpose of proposing an executive summary of financial signals. We show, specifically, that the tendencies we propose are fundamentally different from one that is obtained by a filtering scheme, and because of this, their corresponding summaries will have a different interpretation, and hence, their utility is context-specific.  
% In Section \ref{sec:interpretation} we consider a variety of 
% different real-world signals and interpret their tendency. 
We contrast the low-pass filter with the tendency outcomes, in order to explore how context matters. Interpretability and robustness have a strong bearing on the practical utility of
the proposed tendencies.  The example applications, additionally,  confirm that  the selection criteria for the two proposed tendencies is reasonably robust, computationally.  

One often thinks of a trend as a good executive summary of a signal, and most trends have a rigorous basis, which endows them with consequent interpretative characteristics. A trend, in that case, could be used to `explain' some aspect of the time series or the mechanism that produced the series. 
The types of signals we wish to focus on are finite, possibly  defined at irregular time intervals. Signals for which the determination of a trend is particularly challenging, are those for which there is no knowledge of the context in which this signal arises.

The use of a decomposition as a means to find a trend is not new. For (linear) vector spaces a trend could be found by some criteria that identifies certain combinations of elements in subspaces as more important than others.
 A more traditional way to 
produce a trend, in the sense of an executive summary, is to use filtering.
 In econometrics we find that the 
 Hodrik-Prescott (HP) filter \citep[see][]{hodrick1997postwar} is  used to find the large-scale trend of complex financial data by extracting aspects of market moves that are significant because they are systematic. Given a time series $(Y(i))$, the trend component $(H(i))$ using the HP filter is defined as:
\[
\min_H \left\{ \sum_{i=1}^{N} \left[ Y(i) - H(i) \right]^2 + \lambda \sum_{i=2}^{N-1} \left[ \left( H(i+1) - H(i) \right) - \left( H(i) - H(i-1) \right) \right]^2 \right\}
\]
where $\lambda$ is a free (tunable) smoothing  parameter. In short, this filter finds a signal $H$ which seeks to minimize both the distance to the original series $Y$, and its own second order difference.
This tendency extraction process can be seen as the application of a windowed low-pass filter. The corresponding transfer function is
\begin{equation}
\frac{\hat H}{\hat Y} =  \frac{4 \lambda (1-\cos \omega)^2}{1+4\lambda  (1-\cos \omega)^2},
\label{eq:hp}
\end{equation}
where $\hat H$ and $\hat Y$ are the Fourier transform of the filter output and the input data $Y(i)$, and $\lambda$ is a free (tunable)  parameter.  (The data is assumed to be equally spaced). This filter is capable of handling data from a non-stationary stochastic process, however, it is hard to make sense of its outcome if the time series is not at least $I(2)$ (non-stationary and must be differenced twice to obtain stationarity).
Another way to obtain an executive summary might be by estimating the smoother \cite{smoothingbook} of the time series. The traditional smoother minimizes the variance of a proposed model against data and can be used as a tendency of a time series; its explainability rests upon the validity of the smoother model, and its quality, on how small the resulting variance is. 

A more empirical approach to defining something that might be considered a tendency, is described in \cite{wu2007trend}, wherein they use the Empirical Model Decomposition (EMD). This approach is an adaptive iterative decomposition similar to the ITD we use in this study, however, we use a radically different criterion than they do in determining which decomposition step yields a tendency. The ITD was chosen because it is a decomposition that is fully adaptive (as is the EMD),  and it is aware of  {\it both } the abscissa and the ordinate values and frequencies in the decomposition. We also found it to be very stable numerically and, thanks to some analysis in \cite{itd}, understood to a certain extent as an iterative application of a diffusion map.  What we find most compelling about using the ITD decomposition to find a tendency is that the decomposition takes into account the histogram of what is being reported, and in addition, the extremas in the signal independent of their associated frequency, which leads to insights on potential 
mechanisms associated with the signal generation itself. This is particularly relevant when the time series happens to have a mechanistic origin.

% In this paper we will also use an ITD decomposition to find the tendency.
% The first attempt at defining  a procedure for computing the tendency appears in \cite{itd}. 
% We found that the tendency and the manner in which we generated it using ITD was satisfactory, but the criteria for picking out the tendency among the baselines was far too fragile for signal that were short. Moreover, the tendency, found in that way did not posses interpretability and thus made the tendency, so defined, of limited use.

% In this paper we will in fact define several notions of a tendency as well as a significantly more robust way to obtain them. We  retain the ITD decomposition. We found the  ITD attractive as a method for decomposition of a signal because it handles any length signal and  handles unequally-spaced time series.  Most importantly, because it is sensitive to  persistency in the abscissa as well as in the ordinate coordinates of the data. Specifically, in the ordinate, usually time, the ITD is sensitive to extremas in a signal and thus potentially informative regarding
% its intermittency, usually derived by mechanistic aspects of the generation of the series itself. It is also sensitive to the frequency of events in the ordinate. 

\section{Background}
\label{sec:background}

\cite{itd}  focused primarily on elucidating properties of a time series decomposition proposed by \cite{frei-osorio06}, the  Intrinsic Time Decomposition (ITD).  A proposed application of the ITD was to compute a trend to the time series, this trend was denoted as the {\it tendency} in \cite{itd}. The tendency was arrived at by a series of empirical tests, whereupon a certain member of the ITD decomposition is chosen as the tendency of the signal. In this paper we propose a more robust algorithm for finding the tendency, one that most critically, gives the tendency some degree of interpretability.  
The ITD is a purely algorithmic, non-lossy 
iterative decomposition  of a time series $\{Y(i)\}_{i=1}^N$ into a set of {\it baselines} $\{B^{j=1:D}(i)\}_{i=1}^N$ and {\it rotations} $\{R^{j=1:D}(i)\}_{i=1}^N$.
The decomposition has the form
\begin{equation}
Y(i) = B^D(i) + \sum_{j=1}^{D}R^j(i),\ i=1,...,N.
\label{itddecomp}
\end{equation}
Baselines and rotations satisfy the relation
\begin{equation}
B^j(i) = B^{j+1}(i) + R^{j+1}(i),\; i=1,...,N; \; j=0,...,D,
\label{relation}
\end{equation}
% The decomposition relates these, at level $j=0,1,..,D$
% \[
% B(i)^{j}=B(i)^{j+1}+R(i)^{j+1}, 
% \]
where $B^0:=Y$. We interpret the ITD as a nonlinear operator $\mathcal{L}$ between subsequent baselines $B^j$ to $B^{j+1}$, and write $B^{j+1} = \mathcal{L}(B^j)$.
To explicitly define $\mathcal{L}$, denote by $\{\tau^j_k\}$, $k=1,2,..,K$ the values of $i$ at which the extrema of $B^j(i)$ occur. The extreme values of $B^j$ are called the {\it knots} and denoted simply by $B^j_k :=B^j(\tau_k)$. (In the event that there 
are several successive data points with the same extremal value, we take $\tau^j_k$ to 
correspond to the time of the rightmost of these extremal values). The  baseline $B^{j+1}$ is  constructed by
a piecewise linear formula: for $i \in (\tau^j_k, \tau^j_{k+1}]$, between successive knots,
\begin{equation}
B^{j+1}(i)  = B^{j+1}_k + \frac{(B^{j+1}_{k+1}-B^{j+1}_k)}{(B^j_{k+1}-B^j_k)}(B^j(i)-B^j_k):= \mathcal{L} B^j(i), 
\label{bi}
\end{equation}
 The formula that generates the knots is
\begin{equation}
B^j_{k}:=B^j(\tau_{k}) = \frac{1}{2} \left[ B^j_{k-1} +   \frac{(\tau^j_{k}-\tau^j_{k-1})}{(\tau^j_{k+1}-\tau^j_{k-1})}(B^j_{k+1}-B^j_{k-1}) \right] + \frac{1}{2}  B^j_{k}.
\label{bk}
\end{equation}
The decomposition ends at $j=D$, which is  when a proper rotation cannot be constructed from its baseline; baseline $B^D(i)$ will only have two knots: $B_{k=1,2}^D$, the two end points. The ITD construction guarantees that  the rotation
\begin{equation}
R^{j+1}(i) = B^{j}(i) - B^{j+1}(i), \quad i=1,2,...,N,
\label{ri}
\end{equation}
is monotonic between adjacent extrema.

A note about boundaries. One could interpret the end points $i=0$ and $i=N$ as extrema, and take the corresponding baselineknots to be averages of the first and last pair of extrema, 
$B^j_1= B^j(1)$,  and $B^j_{K^j}= B^j(N)$:
\label{bcs} \\
\begin{equation}
B^{j+1}_1 = \frac{1}{2} (B^j_2+B^j_1)  \quad \mbox{ \ \   and } \quad
 B^{j+1}_{K^j} = \frac{1}{2} (B^j_{K^{j-1}}+B_{K^j}). \label{freebc}
\end{equation}
These will be  called  {\it free} boundary conditions. In special conditions, such as when dealing with periodic signals, we will use \emph{periodic} boundary conditions, meaning we average the first and last point amongst themselves.

\medskip

\begin{figure}[h]
    \centering
    \includegraphics[scale=.7]{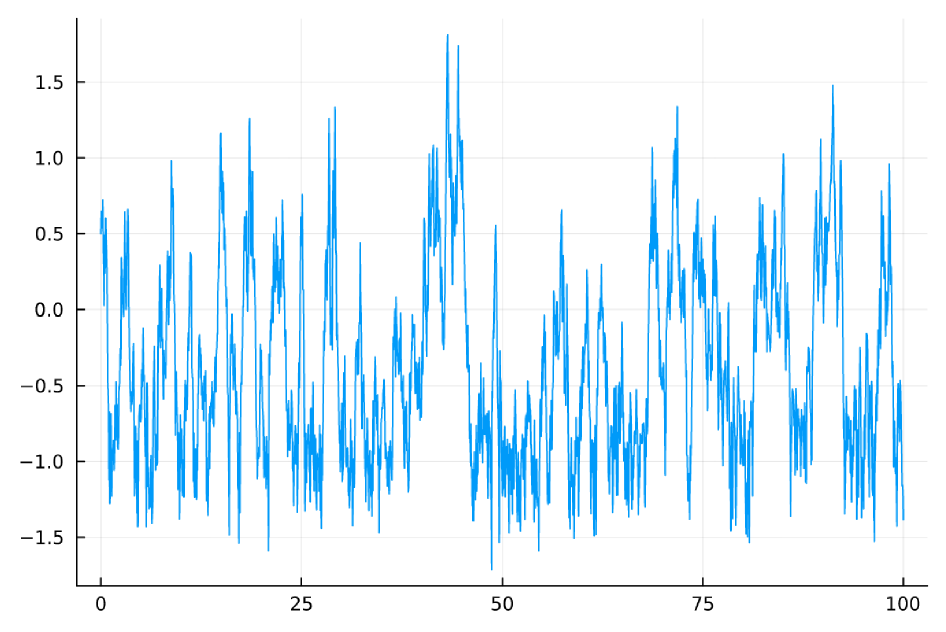}
    \caption{A times series generated by the SDE in \eqref{ex_SDE}. The solution of the SDE was approximated by the Euler-Maruyama method. Here $t_i = i \Delta t$, $i=0:N=2000$ with $\Delta t = 0.05$. }
    \label{f:SDE_time_series}
\end{figure}

\begin{figure}[h]
     \begin{subfigure}{6.7cm}
         \includegraphics[width=\textwidth]{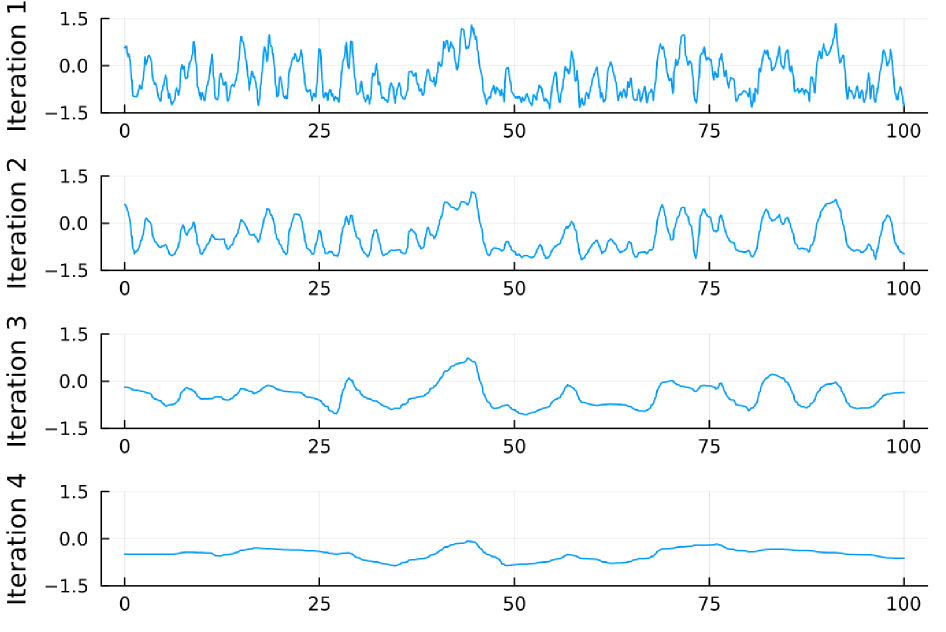}
     \end{subfigure}
     \begin{subfigure}{6.7cm}
         \centering 
         \includegraphics[width=\textwidth]{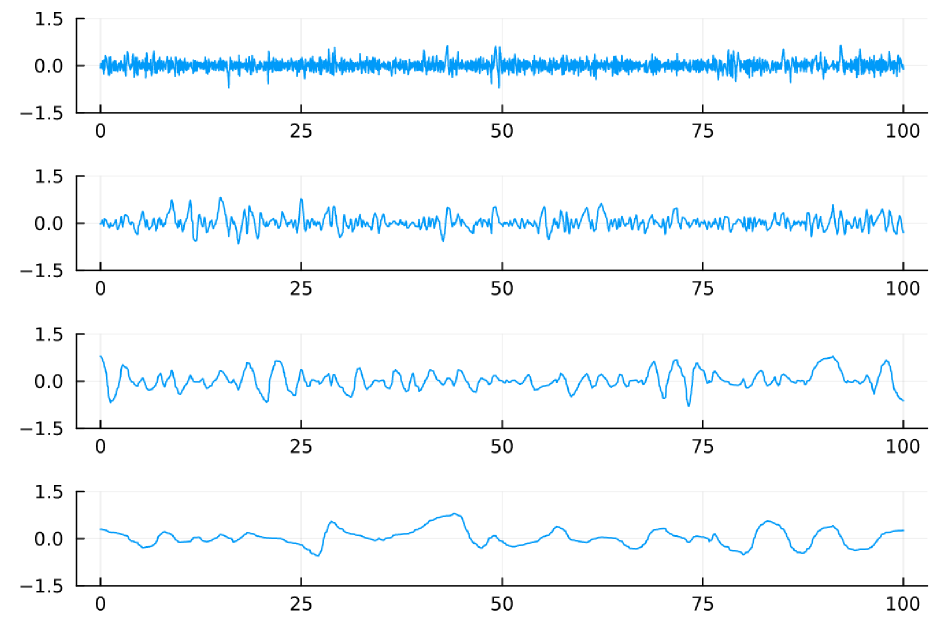}
     \end{subfigure}
    \begin{subfigure}{6.7cm}
         \includegraphics[width=\textwidth]{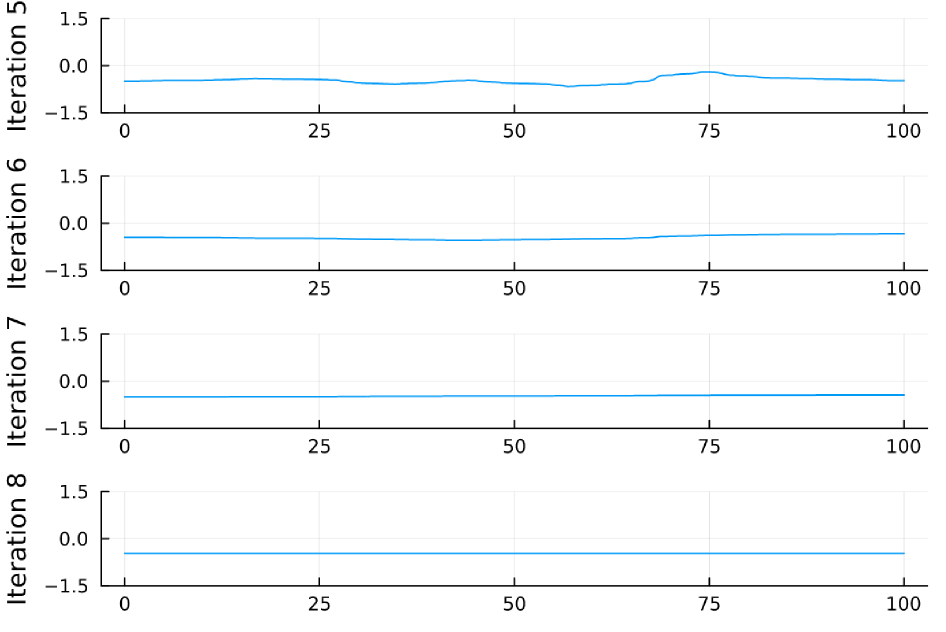}
    \caption{}
     \end{subfigure}
     \begin{subfigure}{6.7cm}
         \centering 
         \includegraphics[width=\textwidth]{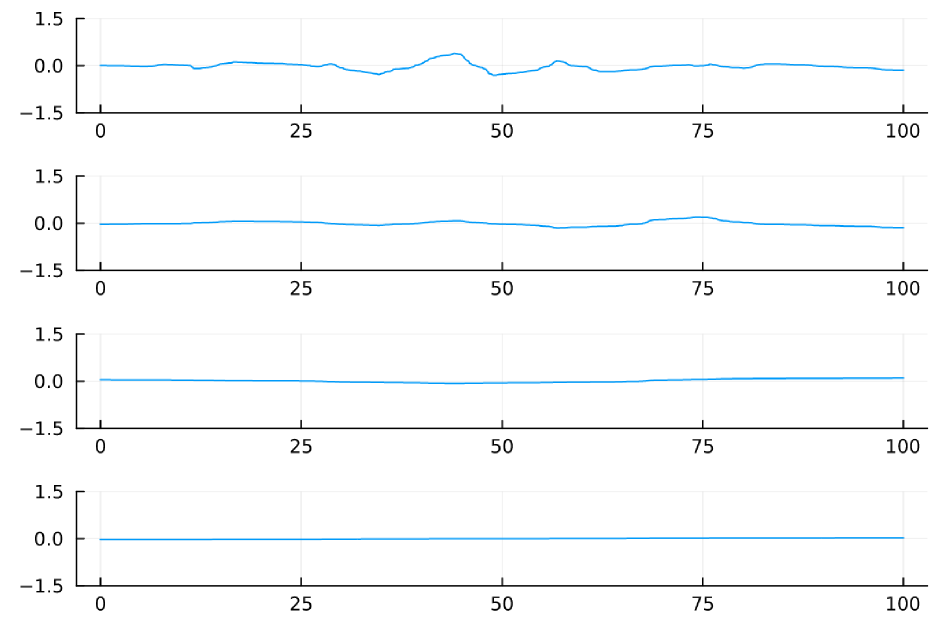}
    \caption{}
     \end{subfigure}
         \caption{Successive (a) baselines, (b) and rotations associated with the signal appearing in Figure \ref{f:SDE_time_series}.  }
    \label{f:sde_itd}
\end{figure}

Figure \ref{f:SDE_time_series} depicts   a realization of the  stochastic differential equation (SDE)
\begin{equation}\label{ex_SDE}
\dif Y_t = -(Y_t^5 - 2 Y_t^4 + 3 Y_t^2) \dif t + \dif W_t, \quad t>0,    
\end{equation}
with $Y(i) = Y_{t_i}$, $Y(0)=0.5$. $W_t$ is an incremental Wiener process.
Associated with this signal is its ITD decomposition shown in Figure \ref{f:sde_itd}. We note some salient and general aspects of the ITD: both the baselines and the rotations get smoother as $j$ increases, the number of $j$ levels ($D=8$ in this example) is small, and the decomposition is non-lossy and generally, non-orthogonal. Moreover, as shown in \cite{itd}, the distance between the extremas and the amplitude of the baselines decreases exponentially fast in $j$.

\begin{figure}[h]
    \centering
    \includegraphics[scale=.8]{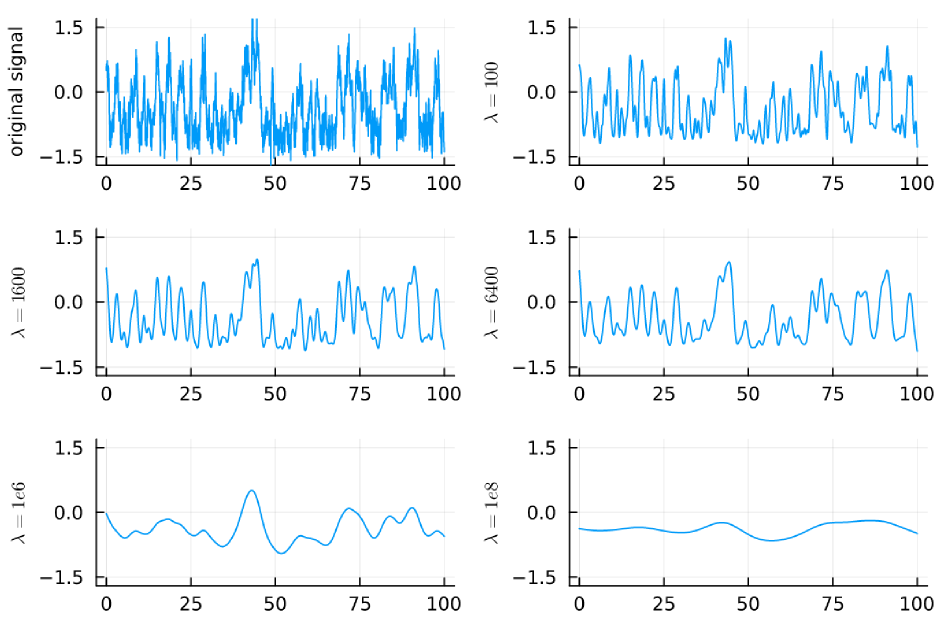}
    \caption{The Hodrick-Prescott filter (see \eqref{eq:hp}), applied to the SDE signal from Figure \ref{f:SDE_time_series}, with different values of the parameter $\lambda$, clearly suggesting that the filter is a windowed low-pass filter. The various outcomes would represent executive summaries of the signal. The choice of  the appropriate $\lambda$ is relegated to the user or is inspired provenance of the data. }
    \label{f:HPSDE_time_series}
\end{figure}

In what follows we will propose the tendency as a particular baseline from the ITD decomposition. and we would like to highlight in what way a tendency is different from a windowed low-pass filter alternative. We will choose the HP filter for comparison. See \cite{hodrick1997postwar}. In Figure \ref{f:HPSDE_time_series} we illustrate the HP filter outcome on the SDE realization of Figure \ref{f:SDE_time_series}, for various values of the filter parameter $\lambda$. In typical applications of the HP to financial data we see $\lambda=1600$ often used, nevertheless, there needs to be some sort of understanding of what to expect from the executive summary when choosing the parameter value $\lambda$.  

%%%%%%%%%%%%%%%%%%%%%%%%%%%%%%%%%%%%%%%%%%
\section{Tendency Selection Criteria}
\label{sec:tendency}

In \cite{itd} a proposal was made for using an information-theoretical metric to pick out the baseline in an ITD decomposition of a time series. 
Namely, it was proposed that a comparison of the  Helinger metric of the baselines could be used to determine the tendency; a significant drop in the metric between two adjacent baselines could used as a criteria. However, this criteria had several flaws, among them, that it was too fragile if the time series was not long enough to generate  rich empirical distributions. Here we propose two new, more robust, criteria for tendency selection in adaptive filters with diffusive behavior. The first, denoted by the Stationary Test Criterion ($\stc$) tendency, inspired by the work of \cite{phillips2019boosting}, selects the ITD baseline by testing the rotations for stationarity.  
The second criteria is algorithmic and is inspired by the diffusion map filtering characteristics of the ITD decomposition itself.  We denote this one the Maximum Extrema Prominence ($\maxep$) tendency. In both cases our criteria selects a step $j^*$ within the ITD decomposition of a signal so that the tendency is defined as $T=B^{j^*}$. Specifically, we propose a decomposition of the form
\begin{equation}
    Y=T+r =B^{j^*} + \sum_{j < j^*} R^j
\end{equation}
where $T$ is the tendency and $r$ is the residual.

\subsection{The Tendency, Based upon the Stationarity Test Criterion ($\stc$)}

The overarching goal of tendency extraction can be interpreted as obtaining a series $T$ that contains known physics and perhaps the epistemic component of $Y$, whilst the residual $r$ will (mostly) contain a largely aleatoric component. Addressing this issue specifically, our $\stc$ criteria tests the rotations at each step $j$, looking to see which $j$ value is the last for which we can say with confidence that the associated rotation is mostly statistically stationary. For stationarity we use the Augmented Dickey-Fuller (ADF) test, which is a widely recognized statistical method to determine the presence of unit root in a given time series sample. The central hypothesis of the ADF test is that a unit root is present in an autoregressive model. The absence of a unit root indicates the time series is stationary or has no time-dependent structure  \citep{cheung1995lag}. 

Formally, consider a basic autoregressive model of order $p$:
\begin{equation}
    \Delta Y(i) = \alpha + \beta i + \gamma Y(i-1) + \delta_1 \Delta Y(i-1) + \dots + \delta_p \Delta Y(i-p+1) + \epsilon(i)
\end{equation}
where \( \Delta Y(i):=Y(i)-Y(i-1) \) is the difference between the current and previous values of the series $Y = (Y(i))$, and \( \epsilon(i) \) is the white noise error term. The ADF test postulates the null hypothesis \( H_0: \gamma = 0 \) against the alternative \( H_1: \gamma < 0 \). If \( \gamma \) is found to be significantly different from zero (based on certain critical value cutoffs), the null hypothesis is rejected, implying the series does not have a unit root and is therefore stationary.
We introduce a functional, denoted as \( \stc \), which operates on a time series and yields the $p$-value associated with the ADF test. For convenience, while the order of the test can be specified, it defaults to 1. To align this with our iterative filtering framework, we represent the ADF $p$-value for the rotation of  series $Y$ as \( ((I - \mathcal{L})^j Y) \), where \( I \) stands for the identity operator.

Our strategy then is to pinpoint the first occurrence where a significant leap in the $p$-value is noticed, indicating the onset of non-stationarity in the series. This becomes our criterion for selecting the optimal level of filtration. We write $\stc$ for the functional that extracts the ADF-test $p$-value for a given time series (from $\mathsf{S}$tationarity $\mathsf{T}$est $\mathsf{C}$riterion). In mathematical notation, the selection criteria is
\begin{equation}
    \label{e:exectuive_summary_iteration_def_adf}
    j^*
    	=
    	\inf
    		\Big\{
    		    j \geq 1; \,
    			\stc(((I - \mathcal{L})^{j+1} Y)) 
    			>
    			p^*
    		\Big\},
\end{equation}
where $p^*$ is selected by the user. In this paper, we use the standard threshold $p^* = 0.05$ for our experiments. One can see that in the cases we present here however, the $\stc$ functional changes abruptly from being very close to $0$ to being quite far from $0$ as we progress through ITD iterations baselines. The choices of the baseline selected by the $\stc$ criterion we present here would not change if we had selected $p^*=0.17$. This leads credibility to our claim that our method does not rely on user-selected parameters.

\begin{figure}[h]
    \centering
    \includegraphics[scale=.7]{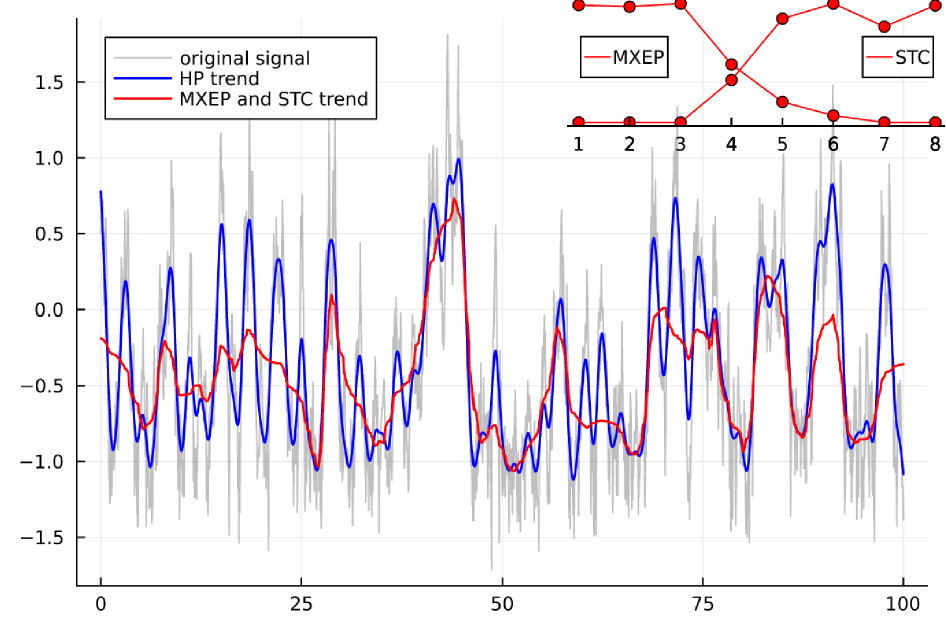}
    \caption{SDE signal (grey) and executive summaries. In blue we depict the outcome of applying the HP filter ($\lambda = 1600$). In red, the ITD-generated
 tendencies, according to $\maxep$ and $\stc$ criteria. Both of these chose the same baseline, in this case, the $j^*=3$. Inset: the 2 criteria, as a function of $j$, on the horizontal axis. The crossing in both cases is at $j=3$, see baseline decomposition of signal on the left row of Figure \ref{f:sde_itd}.  }
    \label{f:baseline_chosen}
\end{figure}

\begin{figure}[h]
    \centering
    \includegraphics[scale=.8]{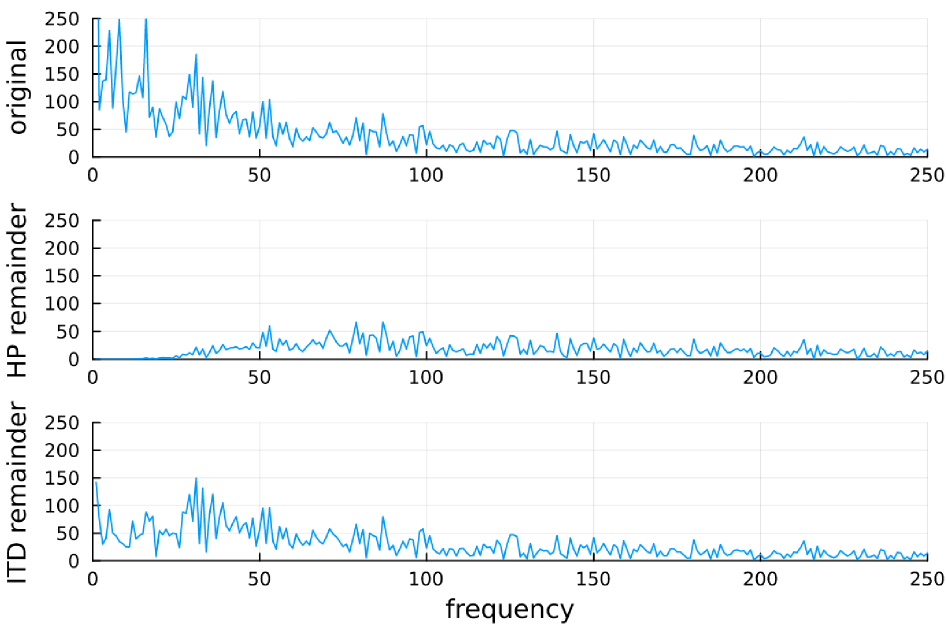}
    \caption{Modulus of the discrete Fourier transform of: the SDE signal, the remainder $r(i)$ obtained by subtracting the HP filtered signal from $Y(i)$ with $\lambda = 1600$; and the remainder $r(i)$ using the ITD tendency for selected $j^*=3$ baseline. Only frequencies with moduli less than 250 are shown.  }
    \label{f:sde_dft_original_hp_itd3_remainder}
\end{figure}

Application of the $\stc$ criteria to the ITD decomposition of the SDE signal identified the $j^*=3$. The baseline $B^3$ is then assigned to be the tendency $T$, and the key difference between the tendency and the data is the removal of 3 rotations, identified as having a statistically near stationary character. The $\stc$ tendency for the SDE signal appears in Figure \ref{f:baseline_chosen}, in red. The inset shows the $p-$value as a function of $j$, it drops after $j=3:=j^*$. Superimposed, in Figure \ref{f:baseline_chosen}, is the HP filter executive summary of the signal. A value of $\lambda=1600$ was chosen for this purpose. Both the HP and the $\stc$ tendency are unlike each other. They are also different in what they filter. As we stated, the $\stc$ tendency filters stationarity in the rotations. The HP filter, on the other hand, is a windowed low-pass filter. In Figure \ref{f:sde_dft_original_hp_itd3_remainder} we make evident this distinction insofar as the spectra of their respective residuals. We note that both techniques produce residuals with high-frequency spectrum essentially equal to the original series, but with important differences in the very-low frequencies. The HP filter acts a pure low-pass filter completely excluding from the residual frequencies below a well-defined threshold. The residual from the ITD decomposition, on the other hand, includes components of every low frequency indicating that a tendency obtained from this method is not reducible to a simple filtering operation. This is due to the fact that at any step of the ITD decomposition, there is the possibility of averaging out variability of any frequency.  
    
\subsection{The Tendency, Based upon the Maximum Extrema Prominence ($\maxep$) Criteria}

To motivate our second criteria, we consider an example of a signal where the difference between neighboring extrema belong to distinct scales. We show how the ITD is only able to average over larger scales after averaging out the smaller ones. The signal we consider is built from a triangular array of $\mathsf{Uniform}[0,1]$ random variables 
\[
    \Big(
        U(k, j);\,
        k=1,\dots,3,\, 
        j=1,\dots,10^{(3-k)}
    \Big).
\]
Then for $i = 0, \dots, 999$, we let the signal be defined as 
\begin{equation}
    \label{eq:multiscale_signal_def}
    \mathsf{multiscale}(i) = 
        100 \cdot
            \sum_{k=1}^3 10^{-(3-k)} 
                \cdot U(k, i \mod 10^i).
\end{equation}
\begin{figure}[h]
    \centering
    \includegraphics[scale=.7]{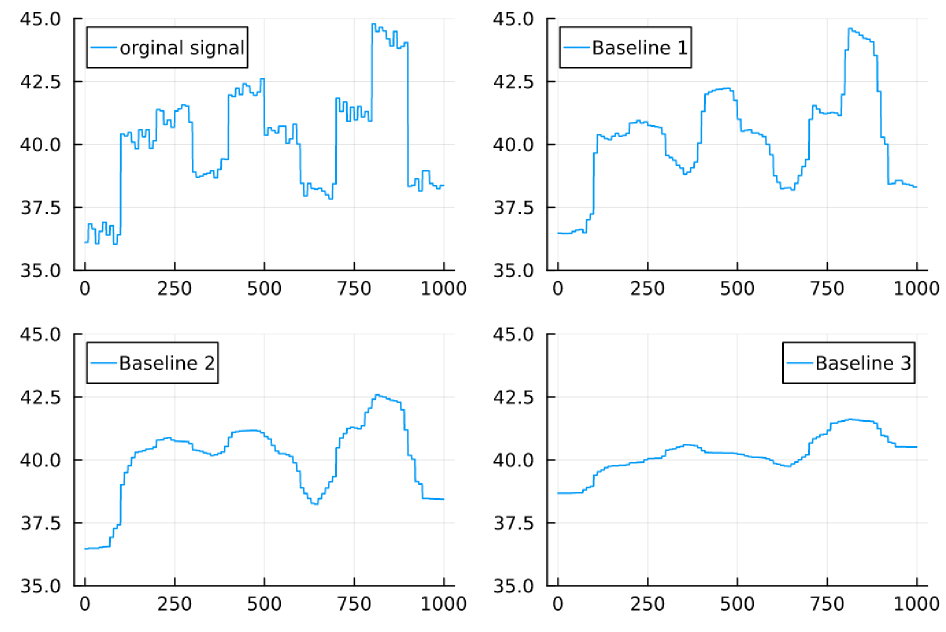}
    \caption{The ITD baselines computed from the multiscale signal defined in \eqref{eq:multiscale_signal_def}.}
    \label{f:multiscale_itds}
\end{figure}
This signal has three distinct scales corresponding to $k=1,2,3$; its plot can be seen in the top-left corner of Figure \ref{f:multiscale_itds}. The rest of the Figure shows the baselines from successive iterations of the ITD algorithm. The ITD works by averaging between local extrema, and at the finest scale, most extrema of the original signal have neighbors whose signal value is within distance $1$. As previously mentioned, only after averaging over the extrema within a scale, is the ITD able to average out extrema of coarser scales: neighboring extrema of the same scale ``protect'' a given extremum from being significantly affected by the ITD. In particular, we see how a prominent valley is erased between iterations $2$ and $3$.

With this multiscale extrema erasure mechanism in mind, we propose a process that is able to discern when the characteristic large scale features of the signal are erased by the smoothing effect of the ITD operator $\LL$. It can be seen as a \emph{maxmin} version of the total variation of a time series, which we call $\maxep$, the \emph{maximum extrema prominence}.

Given an extremum $(\tau_{k_0}, Y(\tau_{k_0}))$ of a time series $Y=(Y(i))$, we consider the neighboring extrema $(\tau_{k_0 - 1}, Y(\tau_{k_0 - 1}))$ and $(\tau_{k_0 + 1}, Y(\tau_{k_0 + 1}))$. They must be of different type then the extrema at $\tau_0$. We can then take
\begin{equation}
    \label{e:prominence_def}
    \prominenceof{\tau_{k_0}, Y}
    	=
    	\min 
    		\Big\{
    			\big|
    				Y(\tau_{k_0}) - Y(\tau_{k_0 - 1})
    			\big|,
    			\big|
    				Y(\tau_{k_0}) - Y(\tau_{k_0 + 1})
    			\big|
    		\Big\},
\end{equation}
as a measure of the (local) prominence of the extremum at $\tau_{k_0}$. It measures how tall of a `peak' or how deep of a `valley' is associated to time $\tau_{k_0}$, by comparing its time-series value to that of its neighbors. We can then define the $\maxep$ associated to the whole time series $Y$ as the maximum of all extrema prominence:
\begin{equation}
    \label{e:maxep_def}
    \maxepof{Y_t}
    	=
    	\max
    		\Big\{
    			\prominenceof{\tau_{k}, Y};
    			\,
    			\tau_k \in \extremaof{Y}
    		\Big\}.
\end{equation}
We note that, had we summed the extrema differences directly instead of extracting this observable through this max-min procedure, we would obtain
\begin{equation}
    \label{e:total_variation_comparison}
    \sum_{\tau_k \in \extremaof{Y}}
        \big|
        	Y(\tau_{k}) - Y(\tau_{k - 1})
        \big|,
\end{equation}
which equals the total variation of $Y$, since the time series in monotone between extrema.

We can now apply this functional to obtain a tendency by applying the process on successive baselines. We obtain a finite sequence of real numbers $(\maxepof{\mathcal{L}^m Y})_{m \ge 0}$.
We then choose the executive summary of $Y$ to be the \emph{last $j$ level} before the value of $\maxepof{\mathcal{L}^j Y}$ \emph{declines the most}. That is, the tendency $T(i)=B^{j^*}(i)$ with 
\begin{equation}
    \label{e:exectuive_summary_iteration_def}
    j^*
    	=
    	\argmin_{j = 0, \dots, D - 1}
    		\Big\{
    			\maxepof{\mathcal{L}^{j + 1} Y}
    			-
    			\maxepof{\mathcal{L}^j Y}
    		\Big\}.
\end{equation}

The heuristic behind tendency selection criterion is that as we iterate the filter, it first smoothens local peaks and valleys, as it only averages between neighboring extrema. However, after microscopic and mesoscopic features are erased, it is finally able to average between extrema whose magnitude is comparable to macroscopic trends. This is when we see the most prominent extrema, captured by the $\maxep$ functional, being erased by the smoothing procedure. So one $j$ level before that happens, we should have an executive summary detailing the macroscopic trends of the time series.

Recalling Figure \ref{f:multiscale_itds} and the multiscale signal $\eqref{eq:multiscale_signal_def}$, we note that the largest prominence drop in seen between iterations $2$ and $3$, and therefore the $\maxep$ criterion chooses $j^*=2$ as the tendency. 
On the SDE signal,  Figure \ref{f:baseline_chosen}, the $\maxep$ tendency is the same as the $\stc$ tendency, the same baseline is chosen. This is not in general the case, as we will see later on.

\section{Interpretation of the Tendencies}
\label{sec:interpretation}

Using a variety of different time series we will highlight important ways in which the executive summary of a time series will differ, depending on which strategy is used in computing the tendency.
 In what follows we contrast the tendency, as computed via the HP filter, and the 2 tendencies, based upon the ITD decomposition, featured in this study.

\begin{figure}[h]
    \centering
    \includegraphics[scale=.7]{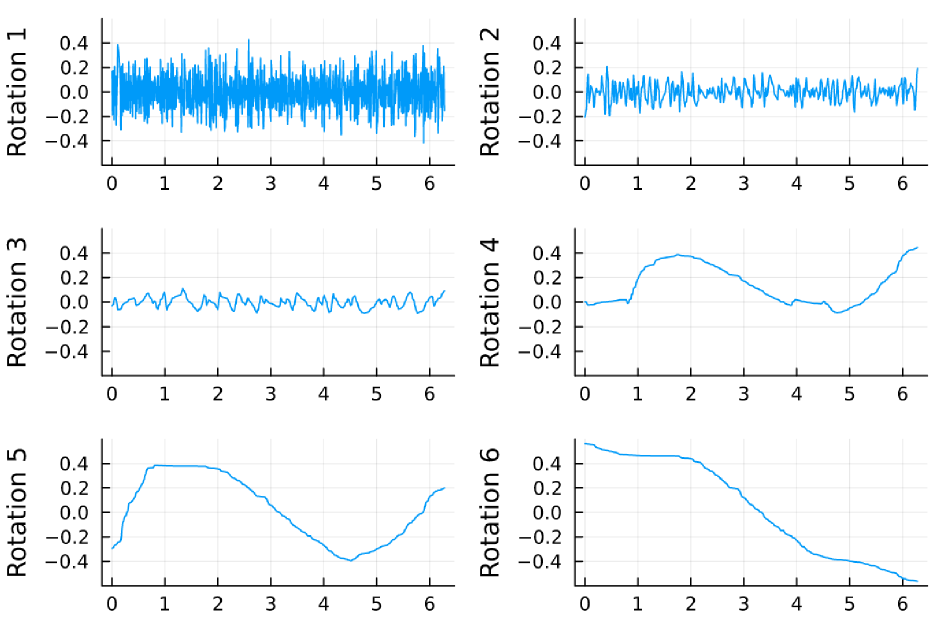}
    \caption{The ITD rotations computed for the noisy sine wave. See Eq.  \ref{eq:noised_sine}.}
    \label{f:noised_sine_rots}
\end{figure}

\paragraph{Noisy sine wave} This is a sine wave in the interval $(0, 2\pi)$ with added normal variate noise:
\begin{equation}
    \label{eq:noised_sine}
    Y(i)=\sin(t_i) + \epsilon_i; 
    \quad t_0 = 0, \, \Delta t = 0.01,
    \, i = \lfloor 200 \pi \rfloor,
    \, \epsilon_i \stackrel{d}{=} \mathcal{N}(0,0.1),
\end{equation}
where $\mathcal{N}(0,0.1)$ is a normal variate with variance $0.1$.  Clearly, this is a signal with significant scale separation.
The expectation is that the HP tendency will be the (low frequency) sine wave itself, given that this executive summary results from the application of a low-pass filter. The outcome, however, depends on how large the $\lambda$ parameter is. The larger it is, the tighter the window.
\begin{figure}[h]
    \centering 
    \includegraphics[scale=.7]{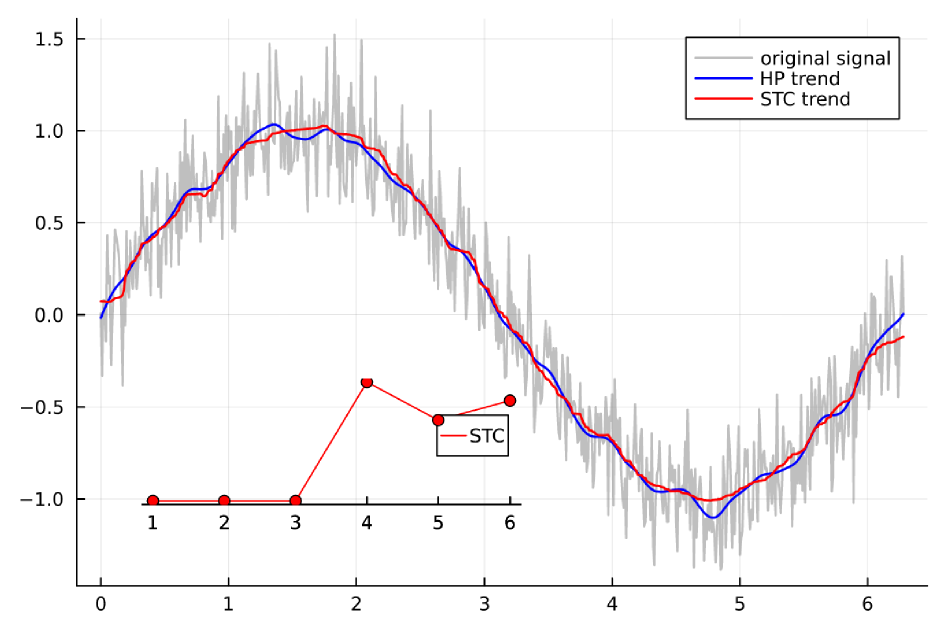}
    \caption{Noisy sine wave together with its tendency based on ITD decomposition algorithm with baseline selected according to the $\stc$ criterion ($j^*=3$). The HP trend is in agreement with the $\stc$-criterion tendency, with $\lambda=1600$.}
    \label{f:noised_sine_trends}
\end{figure}  
The HP trend and ITD $\stc$ tendency are shown  in Figure \ref{f:noised_sine_trends}.
 The baseline chosen by $\stc$ has stripped out rotations that are largely composed of data from a near-stationary random distribution. The full complement of rotations associated with this signal appears in Figure \ref{f:noised_sine_rots}, the large $j$ rotations have a $p$-value that is high, as seen in the inset in Figure \ref{f:noised_sine_trends}.

\paragraph{Time-dependent spectrum time series}

We now consider the time-series defined as 
\begin{equation}
    \label{eq:crazy_signal}
       Y(i) = 10  t_i^3 \cos(13 t_i^3)  \sin(31 \pi t_i) 
\end{equation}
sampled with $t_i \in [0, 2]$ with distance between sampling points $\Delta t = 0.01$. This series exhibits a time-dependent spectrum. Since some of the notable structure is in the high frequencies, we expect the HP trend to deliver less informative results. 
\begin{figure}[h]
\begin{subfigure}{6.7cm}
         \includegraphics[width=\textwidth]{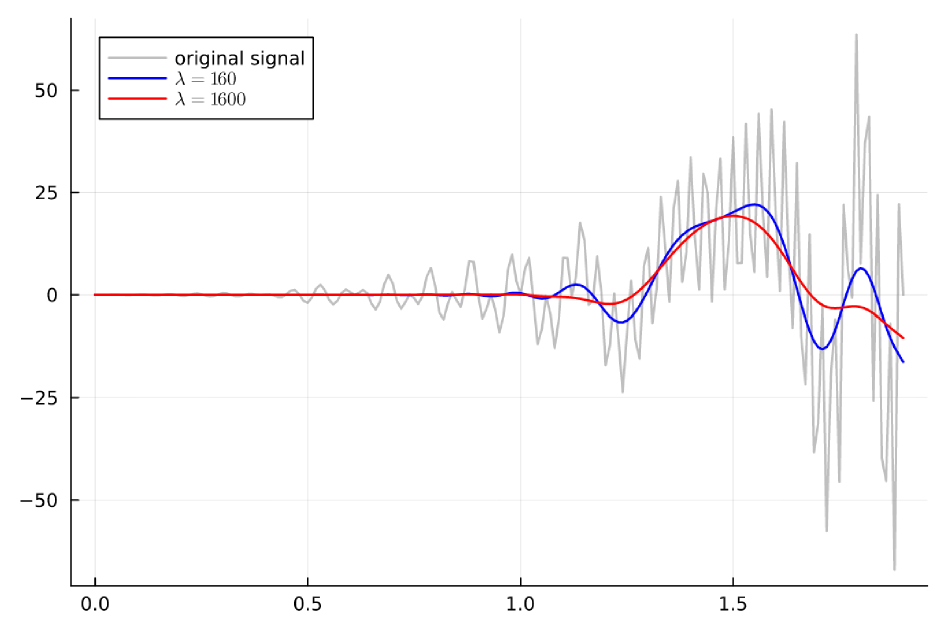}
    \caption{}
     \end{subfigure}
     \begin{subfigure}{6.7cm}
         \centering 
         \includegraphics[width=\textwidth]{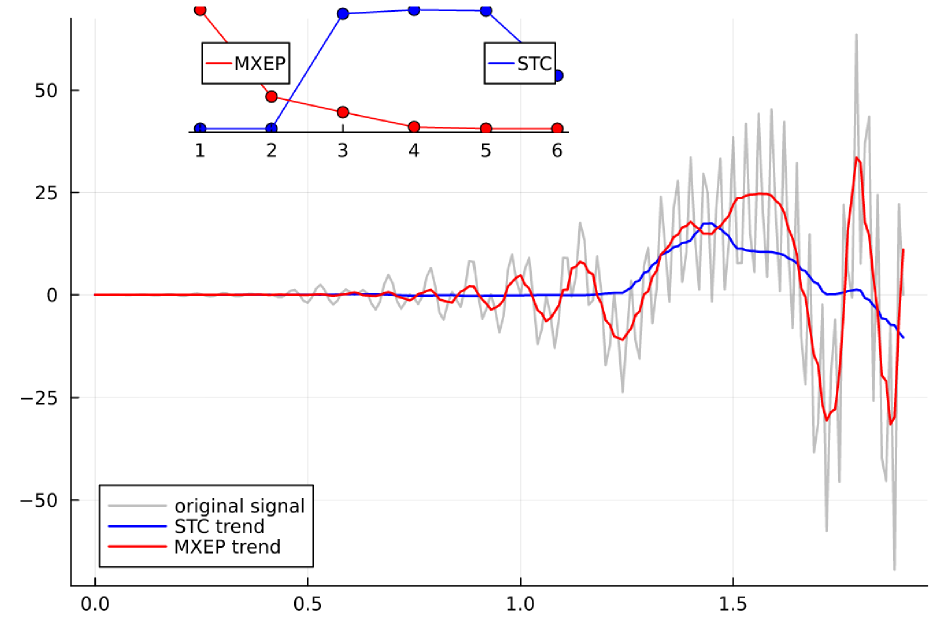}
    \caption{}
     \end{subfigure}
         \caption{Time dependent spectrum time series (gray), \eqref{eq:crazy_signal}. (a) HP filter trend ($\lambda$ small, blue, and $\lambda$ large, red);
         (b) ITD tendencies, with baseline ($j^*=2$) selected according to the $\stc$  and ($j^*=1$) with $\maxep$  criteria. Inset, $p-$values and maximum promimences.}
    \label{f:multispectrum_trends}
\end{figure}
The signal, along with the HP trend and the two ITD tendencies are shown in Figure \ref{f:multispectrum_trends}. The $\maxep$ criterion only eliminates one rotation from the signal, as there is steep decline in the maximum prominence between the first to baselines. The $\stc$, on the other hand, removed the first two rotations, as it did not find strong stationarity in the rest.
It is not possible to favor one tendency over the other. The case for which tendency to choose here 
will have to be decided by considering the 
the signal's provenance. With that said, the tendency that follows the signal the closest is given by the $\maxep$ criterion.

\begin{figure}[h]
\begin{subfigure}{6.7cm}
         \includegraphics[width=\textwidth]{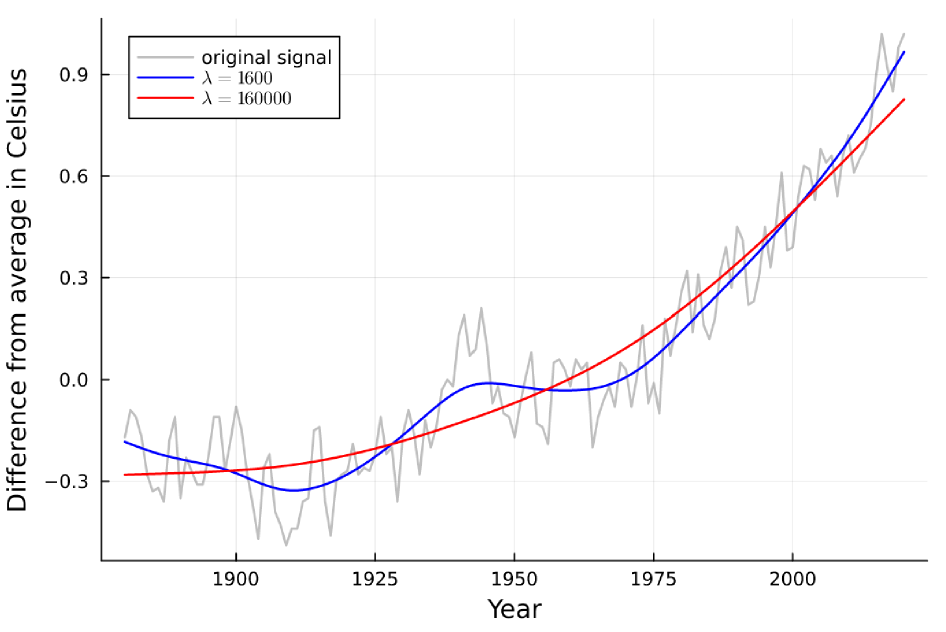}
    \caption{}
     \end{subfigure}
     \begin{subfigure}{6.7cm}
         \centering 
         \includegraphics[width=\textwidth]{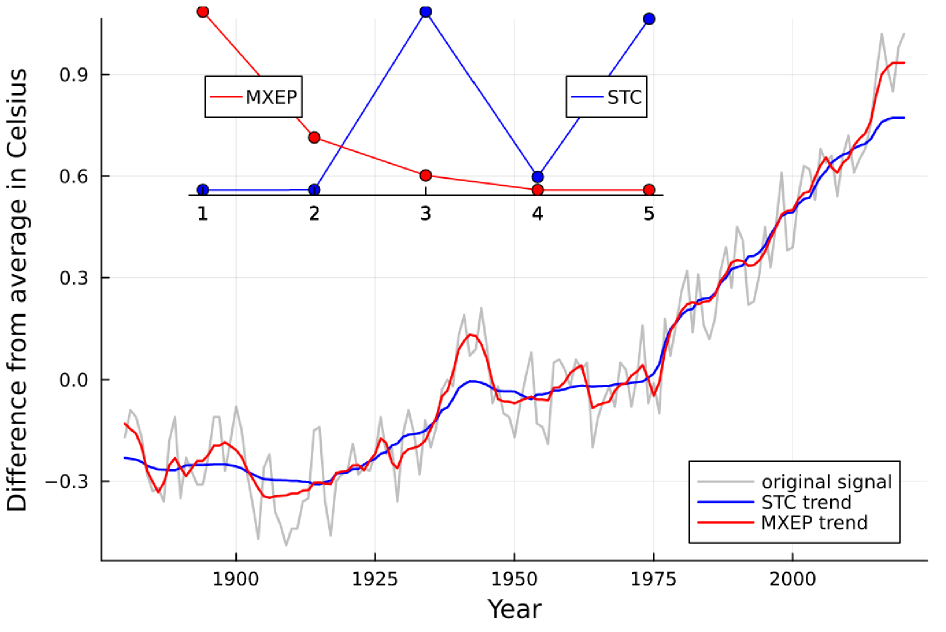}
    \caption{}
     \end{subfigure}
         \caption{(a) The tendency from the ocean temperature anomaly series extracted according to the HP filter ($\lambda=1600$ and $\lambda=160000$); (b) ITD based tendencies. The $\stc$ criteria chose   $\stc$ $j^*=2$ (blue) and $\maxep$ criteria $j^*=1$ (red).}
    \label{f:ocna_trends}
\end{figure}

In the following two examples we consider real-world signals. In  \cite{comou_rahmstorf2011increase}
the authors compute a global temperature for the  post-industrial era. The computation of the trend for this particular data was achieved using a windowing procedure, due to the inherent multiscale spectrum of the signal.  Their trend has a certain degree of explainability because the time window selected for their  analysis is based by the estimated relaxation time for temperatures in the upper ocean, yielding a trend that was consistent with the dynamics and thermodynamics of the system. In Figure \ref{f:ocna_trends}a we display the annual ocean temperature anomaly for over a century (data from GISS), and superimposed, two trends as predicted by the HP filter. The blue curve uses a value of $\lambda$ that yields a tendency in agreement with the results from  \cite{comou_rahmstorf2011increase}. This `correct tendency' is, off course, obtained because the filter parameter $\lambda$ can be deduced from the physical context, which might not be the case for general signals.

In Figure \ref{f:ocna_trends} we superimposed on the temperature anomaly the tendencies obtained through the criteria introduced in Section \ref{sec:tendency}. The criteria used in determining these tendencies was agnostic with regard to the provenance of the signal, and in this regard, less tentative regarding how it proposes a trend, as compared to the HP filter. Regarding the difference between the tendencies, we note that the $\maxep$ tendency is more complex than the $\stc$ counterpart, it follows the signal closer.

\begin{figure}[h]
\begin{subfigure}{6.7cm}
         \includegraphics[width=\textwidth]{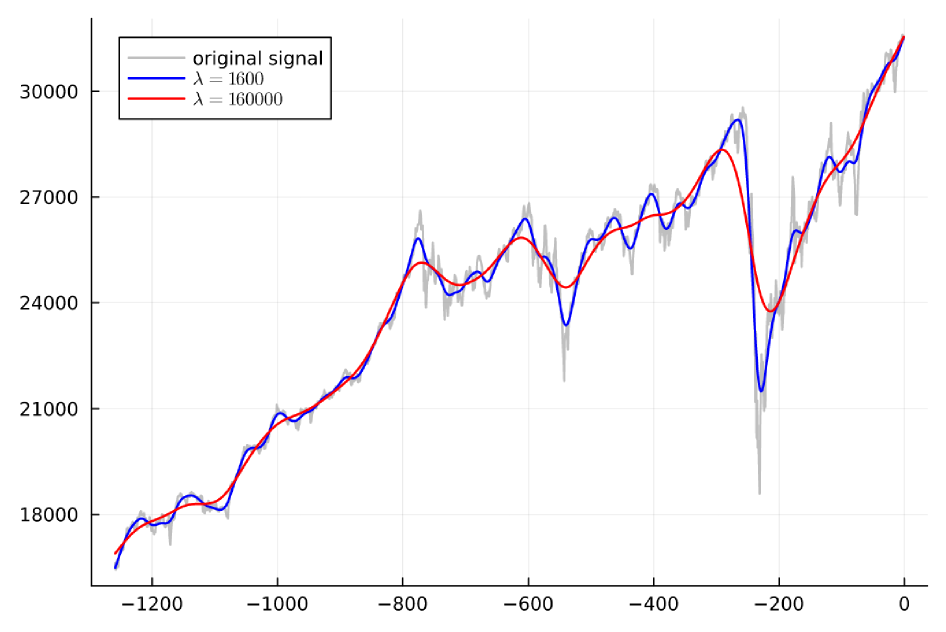}
    \caption{}
     \end{subfigure}
     \begin{subfigure}{6.7cm}
         \centering 
         \includegraphics[width=\textwidth]{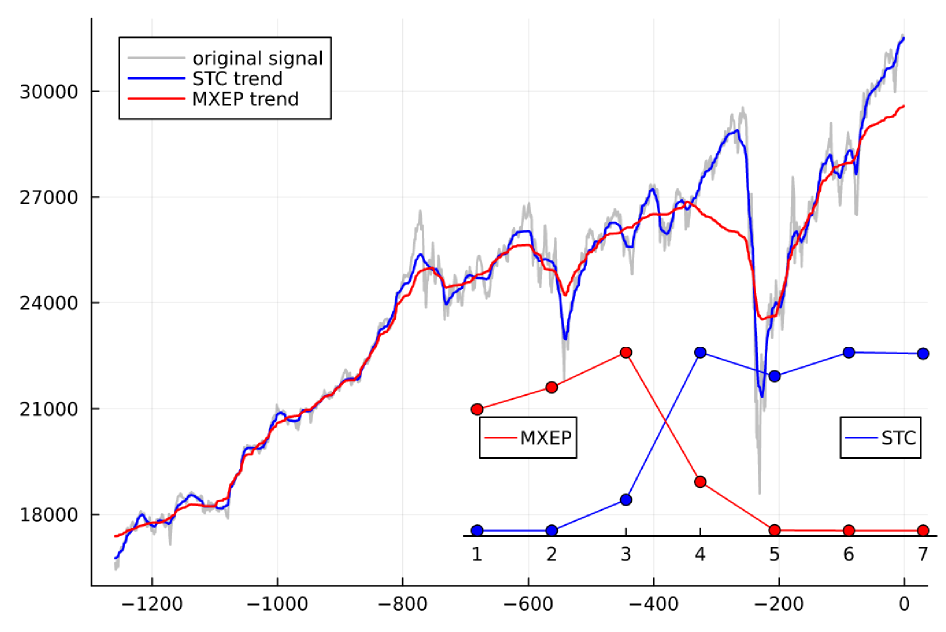}
    \caption{}
     \end{subfigure}
         \caption{The tendency from the DJIA for 1200 days until September 14th of 2023 series extracted according to the HP filter ($\lambda=1600$ and $\lambda=160000$) (a), and ITD algorithm with baseline selected according to the $\stc$ ($j^*=2$) and $\maxep$ ($j^*=3$) criteria.}
    \label{f:djia_trends}
\end{figure}

In Figure \ref{f:djia_trends} we analyse a Dow Jones Industrial financial market time series.
For this data set, again, there was no intervention in suggesting a trend with the ITD. The HP filter trend was found by  trial and error.  The $\stc$ tendency  follows the market trend more closely, the $\maxep$ tendency captures most of the market movement with a considerably less complex tendency.

\section{Discussion and Summary}
\label{sec:summary}
We connote as the  tendency of a signal an executive summary of
a time series, capturing its key dominant features whilst ignoring aspects of the signal that could  be easily parametrized with simple statistical models. The tendency should be less complex than the original signal, and ideally, interpretable. Trends (which might be used to capture an executive summary of a time series), may be endowed partially or in full with interpretability in the sense that these often result from an optimization process. 

Quantitative as well as qualitative executive summaries of data are commonly used in the financial industry. With the advent of machine learning methods and classifiers, physics and engineering has also learned to use low dimensional surrogates or approximate descriptions of complex systems as a way to gain insights into these systems. 
The utility of the tendency 
can go beyond its parsimonious character:
In the context of signals arising in physics and engineering applications, for example, 
an executive summary of a signal may be helpful in discerning what aspects of a signal might have a rational origin and which ones can be ignored, possibly because they present themselves as  aleatoric noise.

In this study we propose two  tendencies extraction methods, both of which originate from the same adaptive decomposition of a time series. The tendencies
 are the result of applying two different  criteria  to identify the tendency among the set of time series decomposition. The tendencies were denoted the maximum extrema prominence ($\maxep$), and the stationary test criterion ($\stc$), after the test that identifies each of these. The specific decomposition algorithm is called the Intrinsic Time Decomposition, proposed by
\cite{frei-osorio06}. The basis for this decomposition is the repeated application of a diffusion mapping (see \cite{itd}).
A notable characteristic of this decomposition is that
it not only is aware of the fluctuations of the signal (the abissa), it also takes into account structure in the ordinate or time axis. Specifically, it elevates the role played by extremas and their distribution in time. 

The time series decomposition allows us to write the time series $Y(i) = T(i) + r(i)$,
where $T$ is the tendency and $r$ the residual. The ITD in turn allows us to write $Y(i) = B^{j*} + \sum_{j< j^*} R^j$, where $B^j(i)$ and $R^j(i)$ are the baseline and rotation time series, for levels $j=1,2,...,D$. The $\stc$ tendency ascribes the $j^*$ baseline to $T$ by testing $R^j$ for statistical stationarity. Hence, the $\stc$ tendency can be interpreted as an executive summary that leaves out aspects of the signal that can be parametrized by a stationary process.  The $\maxep$ tendency is more topological in nature. It selects $j^*$ based upon a {\it maxmin} criterium of the extremum, assigning to $T(i)$ the baseline at $j^*$ level such that prominences still are seen as significant. 

These criteria that dictate which of the baselines in the ITD is picked for the tendency, are, for now, to serve as a guide in making a choice of which tendency to use for a given data set. In the future, a more complete quantitative description of the criteria that select the $\stc$ or $\maxep$ tendency will improve these in terms of how they are to be interpreted as executive summaries for a data set. However, the demonstrations presented in this paper already demonstrate that they are useful time series analyses tools, and that they use criteria that make them different than other possible trends that could be used to produce executive summaries of a time series.

\section* {Acknowledgements}
The submitted manuscript has been authored by a contractor of the U.S. Government under Contract No. DE-AC05-00OR22725. Accordingly, the U.S. Government retains a non-exclusive, royalty-free license to publish or reproduce the published form of this contribution, or allow others to do so, for U.S. Government purposes.

\bibliographystyle{plainnat}
\bibliography{references}
\end{document}